\documentclass[%
reprint,
amsmath,amssymb,
aps, 
physrev,
]{revtex4-2}

\usepackage{xcolor}
\usepackage{tikz}
\usepackage{graphicx}
\usepackage{enumitem}
\usepackage{dcolumn}
\usepackage{bm}
\usepackage{mathtools}
\newcommand{\bfeta}
{\boldsymbol{\eta}}

\begin{document}

\preprint{APS/123-QED}

\title{\textbf{
Long-range correlations in a locally constrained exclusion process}
}%

\author{Stefan Grosskinsky}
\affiliation{%
Institute of Mathematics, University of Augsburg, Augsburg, Germany
}%

\author{Gunter Sch\"utz}
\affiliation{Centro de An\'alise Matem\'atica, Geometria e Sistemas Din\^amicos, Departamento de Matem\'atica, Instituto Superior T\'ecnico, Universidade de Lisboa, Av. Rovisco Pais 1, 1049-001 Lisboa, Portugal
}%

\author{Ali Zahra}
\affiliation{%
Laboratoire de Physique et Chimie Th\'eoriques, Université de Lorraine, Nancy, France
}%
\affiliation{Departamento de Matem\'atica, Instituto Superior T\'ecnico, Lisboa, Portugal
}%

\date{\today}

\begin{abstract}
We introduce a novel exclusion process 
with a simple local kinetic constraint that leads to a nonequilibrium steady state with
a remarkable transition between 
a homogeneous phase with short-range correlations and a clustered phase with long-range correlations and spontaneous breaking of translation invariance. The metastable dynamics of particle clusters lead to a coarsening cascade and glassy dynamics, as well as an intriguing faster-is-slower effect, where an increasing asymmetry in the flow direction leads to a decrease of the stationary current.
\end{abstract}

\maketitle

Lattice gas models have long been a cornerstone of statistical physics, providing a minimal yet powerful framework for studying transport phenomena, phase transitions, and non-equilibrium dynamics \cite{Spohn1991,Privman1997}. Even in one dimension, interacting particle models with noisy dynamics have served as testbeds for understanding hydrodynamic behaviour and anomalous transport on conserved mass, energy
and other physical quantities \cite{Lepr16}.
A paradigmatic example of an exactly solvable lattice gas is the Asymmetric Simple Exclusion Process (ASEP) \cite{Derrida1998,Schuetz2001}, widely recognized as a fundamental model for one-dimensional transport processes with short-range interactions and one conservation law, viz. conservation of mass. The ASEP has found applications in diverse contexts, including  ion conduction \cite{katz1984nonequilibrium}, single-file diffusion in porous media \cite{Kukl96} and narrow channels \cite{Wei00}, vehicular
traffic flow \cite{Helbing2001,Scha10}, biological transport processes through ion channels and along macromolecules and microtubuli \cite{chou2011non,Chow24}, and interface growth in the universality class
of the Kardar-Parisi-Zhang equation \cite{Krie10,Spoh17,Ljub19,DeNa23}.

Many stochastic systems, particularly in fields like traffic flow, granular matter, chemical reactions, and biological motion, involve interactions that extend beyond nearest neighbors. 
An eminent example of a lattice gas incorporating such interactions is the Katz-Lebowitz-Spohn (KLS) model \cite{katz1984nonequilibrium}, a driven kinetic Ising model with conserved
magnetization. More recently, the facilitated exclusion process where particles are pushed forward by neighbors \cite{Ross00,basu2009active,gabel2010facilitated,Ayye23,Lei23} has gained
prominence. This process was originally introduced to study absorbing phase transitions with a conserved field. Once the transient states have died out the absorbing domain remains active and the facilitated exclusion process becomes
equivalent to an exclusion process with extended particles 
with short-range stationary correlations \cite{Laka03,shaw2003totally}. 
The hydrodynamic continuity equation was derived heuristically in \cite{Scho04}
and recently proved rigorously \cite{Blon20,Blon21}.
These models fall under a broader class of transport models with local kinetic constraints \cite{helbing1999global,Bert04,Gonc09}.

A notable feature of such one-dimensional conservative driven diffusive 
systems is the absence of stationary long-range correlations, 
as argued already a long time ago on the basis of a perturbative expansion around equilibrium \cite{Garr90}. Long-range correlations arise only
through breaking of translation invariance e.g. by
defects or boundaries. This picture
echoes the well-known absence of phase transitions in one-dimensional equilibrium systems with short-range interactions
and is now widely accepted. 
Indeed, the only one-dimensional driven diffusive systems for which phase transitions have been reported are not covered by the analysis of \cite{Garr90}. They involve either two or more conservation laws where particles can act as obstacles similar to defects that block the motion for the other species individually \cite{Mall96,Lee97,Cant24} or collectively \cite{Lahi97,Evan98,Kafr03,Chak16,Maha20}
or they exhibit long-range interactions through a logarithmic interaction potential \cite{Priy16,Beli24} or induced by conditioning on atypical
dynamics \cite{Bodi05,Leco12,Jack15a,Kare17}.

In this letter, we introduce a novel exclusion process with a minimal kinetic constraint depending on next-nearest neighbors. Specifically, a particle at site~$k$ can hop forward with rate $q\geq 0$ only if the immediately preceding site~$k-1$ is empty, while hopping backward with rate $1$ requires that both sites~$k-1$ and~$k-2$ are unoccupied. As a consequence, whenever two particles occupy adjacent sites, the forward hop for both is blocked. As shown below, this very simple constraint significantly alters both the dynamical and steady-state properties compared to standard exclusion models or the facilitated exclusion process. 
Our model combines several striking features in one spatial dimension with just a single type of particle. Controlled by the asymmetry parameter \( q \), it exhibits a transition between a homogeneous phase with short-range correlations and a clustered phase with metastable stationary dynamics, a coarsening cascade and glassy relaxation dynamics. These are known to occur under kinetic constraints \cite{ritort2003glassy,cancrini2010kinetically,causer2022slow}, but unlike previously studied systems, our model also exhibits a non-trivial nonequilibrium steady state with long-range correlations and negative differential mobility \cite{sellitto2008asymmetric,chatterjee2018negative}, also known as the faster-is-slower effect \cite{helbing2000simulating}. In the half-filled case, the steady state is analytically tractable, and the dynamics are reversible for all values of $q$, in close analogy with ~\cite{causer2022slow}.

The half-filled stationary theory has been developed in greater detail in Ref.~\cite{Zahra2025}, using a complementary and more rigorous approach. Here we also investigate the nonequilibrium regime below half filling, in particular its transport, metastability and relaxation dynamics.

This model is a natural asymmetric generalization of the celebrated Facilitated Totally Asymmetric Exclusion Process (F-TASEP), and thus inherits much of the physical motivation relevant to F-TASEP \cite{Ross00,basu2009active,gabel2010facilitated,Ayye23,Lei23}. 
For a concrete physical picture, the model can be thought of as an idealized description of clogging phenomena in microfluidic porous channels \cite{Dressaire2017}. In such systems, flow arrest typically occurs when particles come into close contact, forming a "bridge" that prevents further motion until cooperative rearrangements or fluctuations relieve the blockage. By contrast, when particles enter the channel individually, with sufficient spacing, they flow freely without clogging. The three-site kinetic constraint in our dynamics essentially captures this mechanism, with the asymmetry parameter $q$ playing the role of the driving force on particles, and strongly influencing the extent of clogging and hence the overall flow, as will be discussed in detail below.

Consider \(N\) particles on a one-dimensional lattice with \(L\) sites and periodic boundary conditions. Each site $k$ can host at most one particle, indicated by the occupation variable $\eta_k\in\{0,1\}$. A complete particle configuration is denoted by $\bfeta =(\eta_1 ,\ldots ,\eta_L )$ with $N=\sum_{i=1}^L \eta_i$. Particles jump randomly according to
\begin{equation}
    010 \xrightleftharpoons[1]{\text{ $q$ }} 001, \quad \mbox{with rate}\quad q\geq 0\ .
\end{equation}

This model first appeared, though remained unexplored, in \cite{helbing1999global} as a limiting case within a broader class of kinetically constrained exclusion processes. 
It can also be viewed as a non-equilibrium, simplified variant of the stochastic version of the quantum Fredkin spin chains introduced in \cite{causer2022slow}.

The special case \( q = 0 \) corresponds to the facilitated totally asymmetric exclusion process (F-TASEP) \cite{basu2009active} after a hole-particle symmetry transformation. 

For densities exceeding half-filling (\(N > L/2\)), the configuration space fragments into multiple ergodic components, leading to a complete suppression of particle flow, similar to F-TASEP. However, unlike F-TASEP, the system does not have transient states, and the active regime at and below half-filling exhibits 
novel behaviour. 

For densities below half-filling (\(N < L/2\)), the dynamics depend on \(q\). When \( q = 1 \), the average current vanishes, and the system behaves diffusively. When \( q < 1\), the system reaches a steady state in polynomial time (\(\sim L^{3/2}\)) 
with a negative stationary current. In contrast, when \( q > 1 \), the steady state is reached very slowly in super-exponential time (\(\sim e^{CL^2 }\)), and the current is positive. Interestingly, in this regime, the current is not solely determined by the local particle density as in the usual hydrodynamic behaviour \cite{Spohn1991} mentioned in the introduction but also depends on the system size and vanishes in the thermodynamic limit.

At half filling, $L=2N$, let
\begin{equation}
h_k=\sum_{i=1}^{k}(1-2\eta_i),\qquad h_0=h_L=0,
\end{equation}
be the height function, with $h_{\min}=\min_k h_k$. A hole gives an
up-step and a particle a down-step. The minimum height is preserved
by the dynamics, as is the parity of the sites attaining it, so the
configuration space splits into exactly two ergodic components
\cite{Zahra2025}. Shifting the origin to a minimum turns the height
path into a Dyck path. The corresponding translation-invariant energy
is its area in units of complete squares (see Fig.~\ref{fig:Energy}),
\begin{equation}
\label{E}
E(\bfeta)=\frac12\sum_{k=1}^{2N}
\bigl(h_k(\bfeta)-h_{\min}(\bfeta)-\tfrac12\bigr).
\end{equation}

\begin{figure}
    \centering
\begin{tikzpicture}[yscale=1,xscale=1,scale = 0.6]
\draw [fill, color=black!0, fill=black!10] (1,1) -- (2,2) -- (3,1) -- (5,3) -- (7,1) -- (6, 0) -- (5,1)--(4,0)--(3,1)--(2,0)--(1,1);
\draw [very thick] (0,0) -- (2,2) -- (3,1) -- (5,3) -- (8,0) -- (9,1) -- (10,0);
\draw [dashed] (0,0) -- (5,5) -- (10,0);
\draw [dashed] (2,0) -- (6,4);
\draw [dashed] (4,0) -- (7,3);
\draw [dashed] (6,0) -- (8,2);
\draw [dashed] (1,1) -- (2,0);
\draw [dashed] (2,2) -- (4,0);
\draw [dashed] (3,3) -- (6,0);
\draw [dashed] (4,4) -- (8,0);
\draw [thick,green] (2,2) -- (3,3) -- (4,2);
\draw [thick,red] (7,1) -- (8,2) -- (9,1);
\draw [->] (2.9,1.4) -- (2.9,2.6);
\draw [<-] (3.1,1.4) -- (3.1,2.6);
\node at (2.7,2) {$q$};
\node at (3.3,2) {$1$};
\draw (0,0) -- (10,0);
\begin{scope}[yshift = -40]
    \draw [thick] (0,0) -- (10,0);
    \foreach \i in {0,...,10}
    {
        \draw (\i,0) -- (\i,0.1);
    }
    \foreach \i in {2,5,6,7,9}
    {
        \node at (\i+0.5,0.5){};
        \draw [very thick, fill] (\i+0.5,0.3) circle (5pt);
    }
\end{scope}
\end{tikzpicture}
\caption{Example of a configuration with $L=10$ and its height
path. A possible update is shown in green, while a forbidden update
(allowed in ASEP) is highlighted in red. The energy $E=4$ equals
the number of shaded complete squares under the shifted path.}
\label{fig:Energy}
\end{figure}

Within each ergodic component, the stationary measure is reversible
for $q>0$ and takes the form~\cite{Zahra2025}
\begin{equation}
\label{invmeasure}
\pi(\bfeta)=\frac{q^{E(\bfeta)}}{Z_N(q)},\qquad
Z_N(q)=\sum_{\bfeta}q^{E(\bfeta)},
\end{equation}
where the sum runs over that component. At $q=1$ the measure is
uniform, with $Z_N(1)=\frac12\binom{2N}{N}$. For $q<1$ it favors
locally alternating configurations, whereas for $q>1$ it is
dominated by configurations with a single block of $N$ particles,
whose maximal energy is $E_{\max}=N(N-1)/2$.
The large-$N$ asymptotic partition function for fixed $q>1$ is
\begin{equation}
\label{zasymp}
Z_N(q)\simeq Nq^{N(N-1)/2}
\prod_{k=1}^{\infty}(1-q^{-k})^{-1}.
\end{equation}
In particular, a specified alternating configuration has $E=0$
and stationary probability $1/Z_N(q)$, which becomes extremely
small for large $N$. This will provide the escape-time estimate for
clusters below half filling. Detailed proofs, a fuller combinatorial
analysis of $Z_N(q)$, and a more rigorous study of the mixed-order
transition at $q=1$ are given in Ref.~\cite{Zahra2025}.

\begin{figure}
\includegraphics[width=0.5\columnwidth]{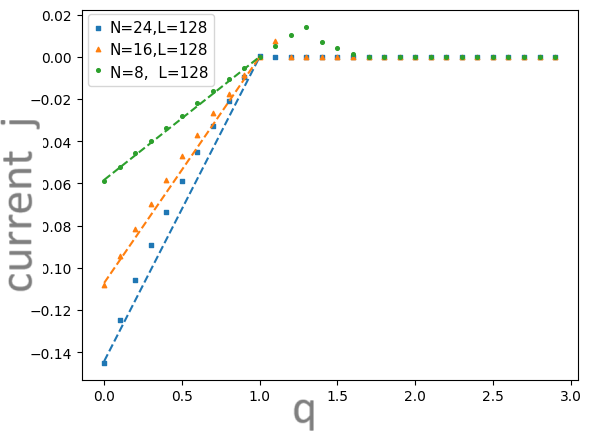}\llap{\raisebox{5mm}
{\includegraphics[height=1.95cm]{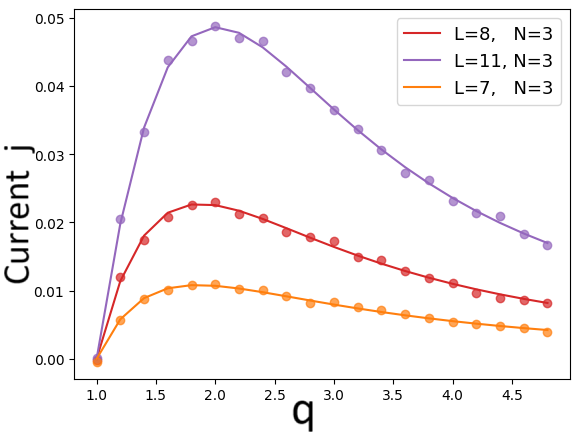}}} 
\raisebox{-1mm}{\includegraphics[width=0.48\columnwidth,height=0.37\columnwidth]{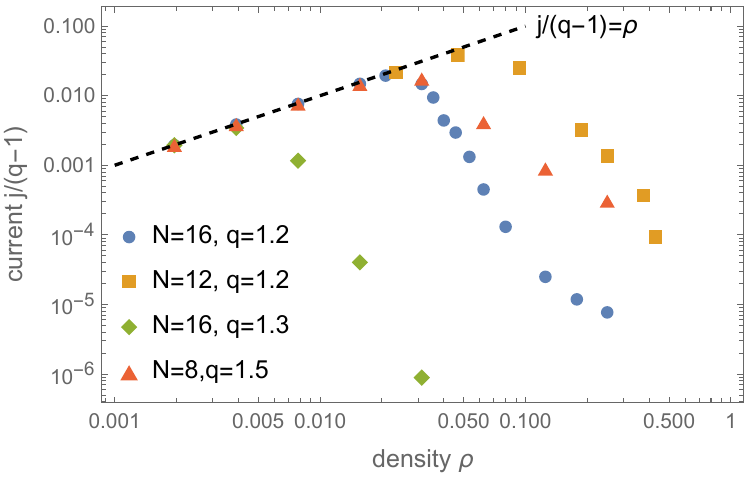}}

\caption{\textbf{Stationary current} as a function of $q$ (left) and $\rho$ (right). 
    For \( q < 1 \), the system exhibits an F-TASEP-like current of $(q-1) J_{F}(1-\rho)$
    (dashed lines on the left). For \( q > 1 \) the current vanishes in the thermodynamic limit with $N/L\to\rho\in (0,1/2)$. Increasing \( q \) leads to a decrease in the current as shown in the left inset for $N=3$ (full lines represent an exact solution) and on the right. For $\rho =0$ with $L\gg Nq^N$ particles can move freely and the current grows linearly as $(q-1)N/L$ (right).
    }
    \label{fig:fd}
\end{figure}

For $L>2N$ it is easy to see that the system has only one ergodic component and we have a unique stationary distribution. This is not an equilibrium distribution for $q\neq 1$ and we have no analytic expression. We expect and observe a non-zero stationary particle current which is positive for $q>1$ and negative for $q<1$. In the latter case, the system relaxes to stationarity on a usual time scale $L^{3/2}$
and its large-scale
relaxation is governed by a standard conservation law with a current
that is a function of the local density. It can be well approximated by $(q-1) J_{F}(1-\rho)$ as shown in Figure \ref{fig:fd}, with $J_{F}(\rho)$ being the facilitated TASEP current given by 
$J_{F}(\rho) = \frac{(1-\rho)(2 \rho - 1)}{\rho}$
\cite{gabel2010facilitated}.

\begin{figure}
\includegraphics[width=0.75\columnwidth]{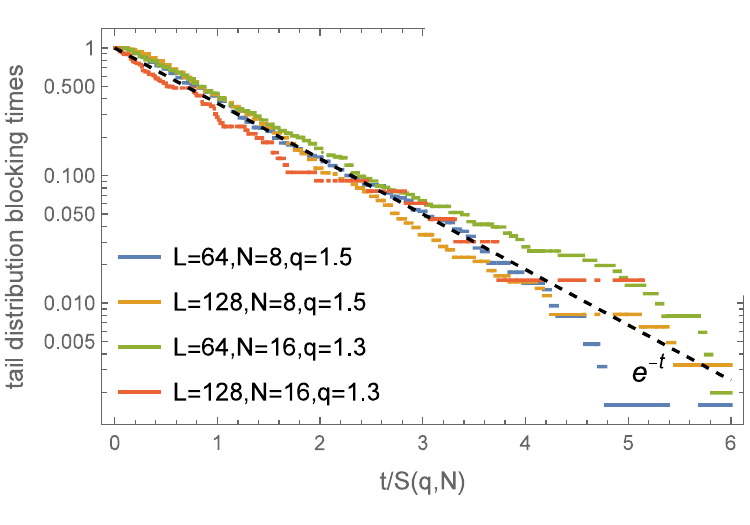}\hfill\llap{\raisebox{23mm}{\includegraphics[width=0.5\columnwidth]{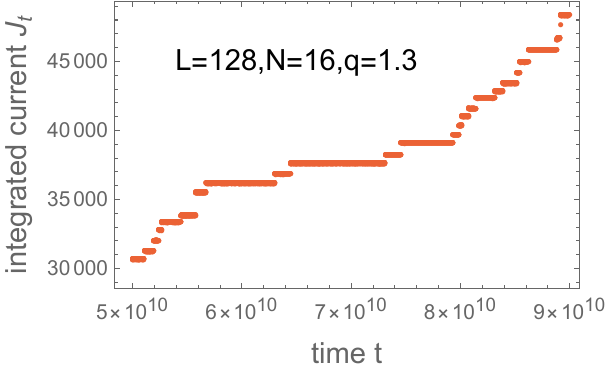}}}
\caption{\label{fig:jumptimes} \textbf{Metastability.} A realization of the time evolution of the integrated current shows abrupt jumps, a characteristic of bottleneck-induced metastable dynamics (top right). The jump times for the stationary current are exponentially distributed on the scale $S(q,N-\frac{C}{1-q})$ (cf.~\eqref{scale}), which is consistent with numerical data (left).}
\end{figure}

For $q>1$ the behaviour of the system is far more intricate, particles drift to the right but  neighboring particles are blocked from jumping to the right. So typical stationary configurations consist of blocks of particles, which have to dissolve before a particle current can flow to the right. To achieve this, a block of $M$ particles has to reach an antiferromagnetic state $\bfeta_M =(0,1,\ldots ,0,1)$, and the expected time for this can be estimated using the stationary probability $1/Z_M (q)$ of such a state in a half-filled subsystem with $2M$ sites. Since the exit rate from $\bfeta_M$ is $q M+1$ and $\pi (\bfeta_M )=1/Z_M (q)$ with \eqref{zasymp} this leads to a time scale
\begin{equation}\label{scale}
S(q,M)=\frac{Z_M (q)}{q M+1} \simeq q^{M(M-1)/2-1}
\end{equation}
Here the reversible half-filled measure is used to estimate the rare unblocking event within a cluster; the full system below half filling is not reversible for $q\neq 1$.
This implies the following features at stationarity:

\noindent $\bullet\ $\textbf{Metastability.} 
Typical stationary configurations are local perturbations of a single block of $N$ particles as extra void does not play a role in this case, in comparison with a half-filled system.
This leads to metastable stationary dynamics on a time scale $S(q,N-\frac{C}{q-1})$ with local corrections on the scale $1/(1-q)$. This can be observed in the step-like time evolution of the stationary particle current, as is demonstrated in Figure \ref{fig:jumptimes}, and indicates spontaneous breaking of translational invariance.

\begin{figure}[t]
\centering
\includegraphics[width=0.55\columnwidth]{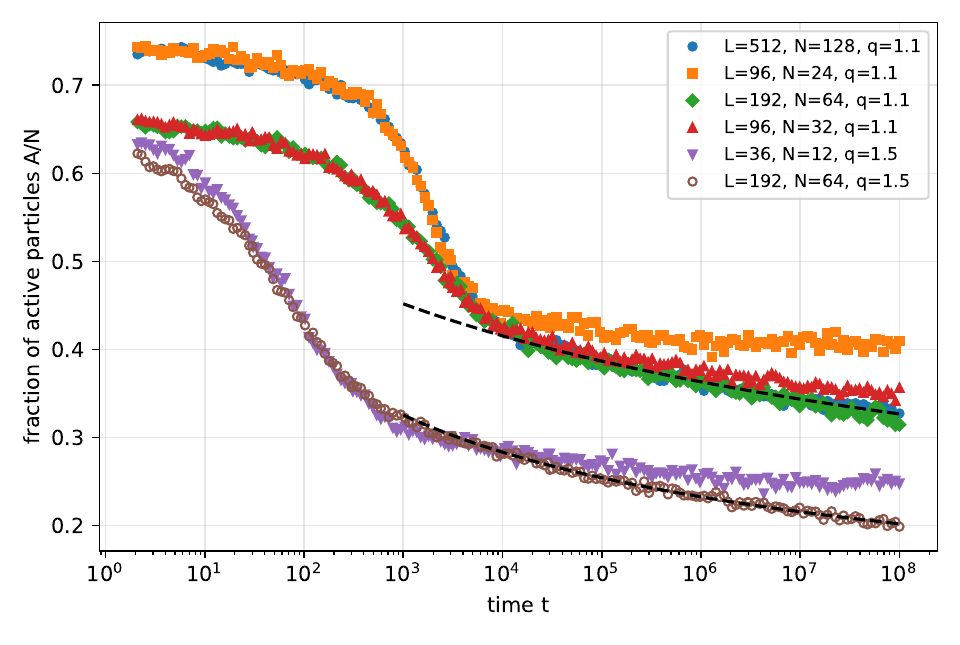}\hfill\raisebox{5mm}{\includegraphics[width=0.43\columnwidth]{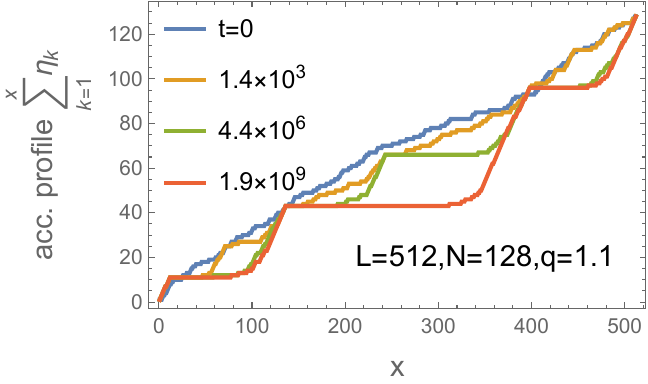}}
\caption{\label{coarsening}\textbf{Coarsening cascades and glassy dynamics.} The density of active particles $a(t)\sim (\ln t)^{-1/2}$ \eqref{aequ} decreases logarithmically slowly and only depends on the parameter $q>1$ after a transient stage, which also depends on the particle density $\rho$ (left;  data averaged over 60 to 400 runs, depending on the system size, dashed lines: fits of $a(t)=(\alpha+\beta\ln t)^{-1/2}$ for $t\geq10^4$). Finite systems deviate once $a(t)$ reaches the stationary value (visible for $L=36$ and $96$). Starting from a uniform initial conditions, accumulated density profiles show coarsening of particle clusters in a typical time evolution (right).}
\end{figure}

\noindent $\bullet\ $\textbf{Negative differential mobility.} An increasing asymmetry (\( q > 1 \)) causes blocking times to grow super-exponentially, leading to a decrease in the stationary current. This counterintuitive effect, also known as faster-is-slower \cite{helbing2000simulating}, occurs when higher individual particle speeds $q-1$ hinder overall transport efficiency. In the thermodynamic limit (\( L, N \to \infty \) with \( N/L \to \rho \)), the current vanishes as $\sim 1/q^{N-1}$ for any density \( \rho \in (0,1/2) \). Only at vanishing densities do particles have sufficient space to move independently, with a crossover occurring at approximately \( L \approx Nq^N \), as illustrated in Figure \ref{fig:fd} (right).
The stationary properties detailed above indicate a breakdown of the standard 
hydrodynamic approach based on a local stationarity and a current-density relation. 
The local kinetic constraint leads to the blocking mechanism explained above, 
which also dominates the dynamics of the model far from equilibrium, 
leading to the following properties:

\noindent $\bullet\ $\textbf{Coarsening.} Metastable clustered configurations exist on all scales and lead to a coarsening cascade when approaching the stationary state from homogeneous initial conditions (see Figure \ref{coarsening} right).
On an infinite system, the coarsening dynamics leads to an ever-increasing average size $m(t)$ of blocked clusters. It is analytically and numerically convenient to quantify the coarsening by the relative density of active particles $a(t)\in [0,1]$ which are not blocked by other particles. We have $a(t)\sim 1/m(t)$ decreasing in time with a rate given by the inverse metastable time scale $S(q,m(t))$, leading to
    \begin{equation}\label{aequ}
        \frac{d}{dt}a(t)\simeq -q^{-a(t)^{-2}/2+1} (1+q)\ .
    \end{equation}

\noindent $\bullet\ $\textbf{Glassy dynamics.} 
For large times the right-hand side of \eqref{aequ} is dominated by the factor $q^{-a(t)^{-2}/2}$, and the solution decays asymptotically as $a(t)\simeq\bigl(\ln q/(2\ln t)\bigr)^{1/2}$, i.e.\ $a(t)\sim (\ln t)^{-1/2}$. This asymptotic regime is approached only very slowly. In the time window accessible to simulations, the data are well described by $a(t)=(\alpha+\beta\ln t)^{-1/2}$ with fitted constants $\alpha$ and $\beta$, see Figure \ref{coarsening} (left). In particular, this decay universally only depends on the parameter $q$ and systems with different particle densities collapse after a transient initial phase. 
In finite systems, the rescaled number of active particles $A(t)/N$ coincides with $a(t)$ very well even for moderate system sizes until the system reaches stationarity.

The constrained ASEP studied in this Letter shows that the
long-standing belief that there is no phase transition in translation invariant driven diffusive systems with one conserved quantity and short-range interactions is incorrect. It exhibits a phase transition from a homogeneous state with short-range correlations to a phase-separated state with long-range
correlations
that exhibits metastability, coarsening, glassy dynamics and a curious faster-is-slower
phenomenon of the stationary current. These properties open up a pathway
to investigating the breakdown of regular hydrodynamics
due to building up long-range correlations
that invalidate the assumption of local equilibrium, which is at the heart
of deriving large-scale hydrodynamic behaviour with a current that is a function of the local density \cite{Spohn1991}.

The breakdown of hydrodynamics also puts into question the nature of stationary fluctuations as one approaches the
transition point. In the hydrodynamic regime $q<1$ they are expected
on general grounds discussed in \cite{Krie10,Spoh17} to belong to the universality class of the 
Kardar-Parisi-Zhang equation with dynamical exponent $z=3/2$ , while at the
transition point similar considerations suggest Edwards-Wilkinson
universality with dynamical exponent $z=2$, both being the lowest 
members of an infinite discrete
series of dynamical universality classes whose dynamical exponents
are given by the ratios of consecutive Fibonacci numbers \cite{Popk15b}.
For $q=1$ the expected Edwards-Wilkinson universality is proved for the facilitated exclusion process in \cite{Erig23} which also gives a bound on the dynamical exponent
in the weakly driven case that is consistent with KPZ scaling. An extension
and sharpening of these results to the present constrained ASEP
and,  more significantly, the nature of the fluctuations near the transition point for $q>1$
is an open problem whose solution would elucidate more generally
the fluctuation patterns at the onset of nonequilibrium phase separation
in kinetically constrained dynamics.

Finally, we point out that the constrained exclusion process sheds new light on the celebrated positive-rates conjecture for cellular automata
\cite{Ligg85}, which proposes that no phase separation in one-dimensional
systems can occur unless certain transition probabilities vanish. Disproving this
conjecture required a two-hundred page proof with a very intricately constructed counterexample \cite{Gacs01}. The present model with ergodic continuous-time dynamics provides a simple counterexample to this
conjecture, since in any discrete-time sampling of the continuous-time dynamics no transition probabilities vanish. 
Hence the role of zero transition probabilities for phase transitions in one
dimension away from thermal equilibrium needs to be reconsidered.

\begin{acknowledgments}
A.Z is grateful to M. Henkel for useful discussions. We thank J.P. Garrahan for pointing out reference \cite{causer2022slow}.
We acknowledge
financial support by ANR-PRME Uniopen, project (ANR-22-CE30-0004-01), FCT (Portugal) through project UIDB/04459/2020 (doi 10-54499/UIDP/04459/2020), and by the grants 2020.03953.CEECIND (doi 10.54499/2020.03953.CEECIND/CP1587) and 2022.09232.PTDC (doi 10.54499/2022.09232.PTDC).
\end{acknowledgments}
\bibliography{apssamp}
\end{document}